\documentclass[]{iopart}

\usepackage{epsfig}
\begin{document}

\title{Vortex shedding and drag in dilute Bose-Einstein condensates}

\author{T. Winiecki$^1$, B. Jackson$^1$, J. F. McCann$^2$, and C. S. Adams$^1$}
\address{$^1$Dept. of Physics, University of Durham, Rochester Building,
South Road, Durham, DH1 3LE, England. UK}
\address{$^2$Dept. of Applied Mathematics and Theoretical Physics, 
Queen's University, Belfast, BT7 1NN, Northern Ireland. UK}

\date{\today}

\begin{abstract}

Above a critical velocity, the dominant mechanism of energy transfer between
a moving object and a dilute Bose-Einstein condensate is vortex formation.
In this paper, we discuss the critical velocity for vortex formation and
the link between vortex shedding and drag in both homogeneous and
inhomogeneous condensates. We find that at supersonic velocities
sound radiation also contributes significantly to the drag force. 

\end{abstract}

\section{Introduction}

One enticing consequence of the discovery of Bose-Einstein condensation
(BEC) in dilute alkali vapours \cite{fermi} is the potential for
refining our understanding of quantum fluids. In particular, the dilute Bose
gas provides a near-ideal testing ground for elucidating the role of vortices 
in the onset of dissipation in superfluids. Recent experiments
have demonstrated the formation of quantized vortices by rotational 
excitation of one \cite{madi00} and 
two-component condensates \cite{matt99}, in analogy with the famous 
`rotating bucket' experiments in liquid helium \cite{donn91}. In addition, 
by moving a far-off resonant laser beam through a condensate, 
Raman {\it et al}. \cite{rama99} measure a heating rate which suggests a 
critical velocity for dissipation characteristic of vortex shedding.

The attractive feature of experiments on dilute Bose gases 
is that the weakly-interacting limit permits quantitative
comparison between theory and experiment. The theoretical
description is based on the Gross-Pitaevskii equation, a form
of non-linear Schr\"odinger equation (NLSE) \cite{ginz58}. In the NLSE model,
the critical velocity for vortex shedding by a cylindrical object was
found to be, $v_{\rm c}\sim 0.4 c$,
where $c$ is the speed of sound \cite{fris92}. In a trapped
(inhomogeneous) condensate the critical velocity is
lower due to the reduction in density, and hence 
sound speed, towards the edge of the condensate \cite{rama99,jack00b}.


In this paper, we study vortex shedding due to the motion of
an object through homogeneous and trapped Bose-Einstein condensates
 using the NLSE model.
The hydrodynamical properties of a NLSE fluid are reviewed in Sec. \ref{sec:nlse}.
In Sec.~\ref{sec:vc} and \ref{sec:drag} we discuss the critical velocity for vortex nucleation and the link between vortex formation and drag in homogeneous condensates.  
In Sec. \ref{sec:trap} we consider the properties of trapped
condensates and highlight the differences from the
homogeneous case.

\section{Quantum fluid mechanics}
\label{sec:nlse}

At low temperatures and low densities, atoms interact by elastic $s$-wave scattering, and  collisions can be parameterised by a single variable, the scattering length, $a$.
For atoms of mass $m$, the wavefunction of the condensate,
$\psi(\mbox{\boldmath $r$},t)$, is given by the
solution of the time-dependent Schr\"odinger equation:
\begin{equation}
i\hbar \partial_t\psi(\mbox{\boldmath $r$},t)
 =  \left[-{\hbar^2 \over 2m}\nabla^2 +
V(\mbox{\boldmath $r$},t)
+ g \vert\psi(\mbox{\boldmath $r$},t)\vert^2\right] \psi(\mbox{\boldmath $r$},t)~,
\label{eq:nlse1}
\end{equation}
where the wavefunction is normalised to the number of atoms, $N$,
the coefficient of the non-linear term, $g=4\pi\hbar^2a/m$, 
describes the interactions within the fluid, and  
$V(\mbox{\boldmath $r$},t)$ represents external potentials arising from the trap 
and any moving obstacle.

\subsection{Fluid equations}

The link between the nonlinear Schr\"odinger equation (\ref{eq:nlse1}) 
and the equivalent equations of fluid mechanics is well known 
\cite{donn91,nozi90}. However some points concerning 
condensate flow near penetrable objects are less familiar. In this section 
we gather some of the key concepts and equations
that figure in the discussion of our simulations.
 
Classical (isentropic) fluid mechanics is based on two coupled
differential equations: one describing the transport of mass, the other the transport
of momentum \cite{land87}.  The relevant quantum variables can be constructed from the wavefunction:
the mass density $\rho$ and momentum current density $\mbox{\boldmath $J$}$ are defined as,
\begin{equation} 
\rho~\equiv~m\psi^*\psi \ \ \ {\rm and} \ \ \
J_k \equiv (\hbar/2i)(\psi^* \partial_k \psi -\psi \partial_k \psi^*)~,
\end{equation}
where the index $k$ denotes the vector component.
The fluid velocity
is defined by $v_k\equiv J_k/\rho$, or equivalently in terms 
of the phase, $\phi$, of the wavefunction, 
$v_k \equiv (\hbar/m)\partial_k \phi$. Clearly, the velocity 
field is a potential flow, however it is also compressible 
and furthermore can support circulation (vorticity) as will be seen.
 
The conservation of mass (probability), i.e., the 
continuity equation, follows from the definition of $\rho$ and 
equation (\ref{eq:nlse1})
\begin{equation}
\partial_t \rho+ \partial_k J_k =0~. 
\label{eq:mass}
\end{equation}
The conservation of momentum equation may be found by
considering the rate of change of the momentum current density,   
\begin{equation}
\partial_t J_k + \partial_j T_{jk}+\rho \partial_k (V/m) =0~,
\label{eq:mom}
\end{equation}
where the momentum flux density tensor takes the form \cite{fris92,wini99a},
\begin{equation}
T_{jk}=\frac{\hbar^2}{4m}(\partial_j\psi^*\partial_k\psi-
\psi^*\partial_j\partial_k\psi+{\rm c.c.})+\frac{g}{2}\delta_{jk}|\psi|^4~.
\end{equation}
This can be rewritten as,
\begin{equation}
T_{jk} =   \rho v_j v_k - \sigma_{jk}~,
\label{eq:momtens}
\end{equation}
where the stress tensor $\sigma_{jk}$ is given by,
\begin{equation}
\sigma_{jk}= -{\textstyle{1 \over 2}}\delta_{jk} g (\rho/m)^2 +(\hbar/2m)^2 \rho 
\partial_j \partial_k \ln \rho~.
\label{eq:shear}
\end{equation}

\subsection{Pressure, sound and drag}

The form of equations (\ref{eq:mass}), (\ref{eq:mom}), and (\ref{eq:momtens}) is identical
to those for classical fluid flow \cite{land87}, the difference emerges from the
nature of the stress, equation (\ref{eq:shear}). A classical ideal
fluid is characterised by $\sigma_{jk}=0,$ for all $ j,k$. 
In a viscous fluid, the shear stress ($\sigma_{jk}, j\neq k$) 
is produced by velocity gradients between neighbouring streams
such that,
$\sigma_{jk}= \eta ( \partial_j v_k + \partial_k v_j)$, where
$\eta$ is the coefficient of viscosity. This creates a frictional force
which gives rise to energy loss. In a pure dilute Bose-Einstein condensate there is no 
frictional viscosity, but a shear stress arises from density gradients, 
the second term in equation (\ref{eq:shear}).
This property gives rise to the possibility of vortex formation and drag without
viscosity.
 
The pressure (normal stress, $-\sigma_{jk}, j=k$) within the 
quantum fluid takes the simple form,
\begin{equation}
p= {\textstyle{1 \over 2}} g (\rho/m)^2 -(\hbar/2m)^2 \rho \nabla^2 \ln \rho~.
\label{eq:pressure}
\end{equation}
The second term, called the quantum pressure, is weak in homogeneous
regions of the fluid, that is, far from obstacles or 
boundaries, vortex lines or shocks.  The essential difference
between interacting and noninteracting (ideal) fluids is
the existence of interaction pressure which supports sound propagation. 
In the bulk of the fluid, where the quantum pressure is negligible,
the speed of sound \cite{nozi90,land87},
\begin{equation}
c=\sqrt{\partial p / \partial \rho} = (g\rho/m^2)^{1 \over 2}~.
\label{eq:sound}
\end{equation}

The force on an obstacle moving through a condensate can be calculated from the rate
of momentum transfer to the fluid. 
By integrating Eq.~(\ref{eq:mom}), one finds that 
the $k$-th component of the force,
\begin{equation}
F_{k}= \partial_t\int_{\Omega} {\rm d}\Omega J_k=-\int_{S} {\rm d}S ~n_j T_{jk}-
\int_{\Omega} {\rm d}\Omega ~ \rho \partial_k (V/m)~,
\label{eq:force} 
\end{equation}
where $S$ is the surface of the object or control surface within the
fluid \cite{wini99a}, $\Omega$ is the volume enclosed by $S$,
$n_j$ is the $j$-component of the normal vector to $S$, and
${\rm d}S$ is a surface element.
The second term on the right-hand side can be likened to the buoyancy
of the fluid.  In the case of homogeneous flow past an
impenetrable object (Sec. \ref{sec:vc} and \ref{sec:drag}), the wavefunction 
vanishes on the object surface and the potential is uniform elsewhere, 
therefore, only the first term contributes. Conversely, for a penetrable
object in a
trapped condensate (Sec. \ref{sec:trap}), $\Omega$ may be chosen to encompass the
entire fluid, and the first term is negligible compared to the second.

\subsection{Quantisation of circulation}

The quantum Euler equation follows from combining the equations
describing the conservation of mass 
and momentum, (\ref{eq:mass}) and (\ref{eq:mom}), along with the identity, 
\begin{equation}
\rho^{-1} \partial_j [ \rho \partial_j \partial_k \ln \rho] =
2 \partial_k [ \rho^{-{1 \over 2}} \partial_j \partial_j \rho^{1 \over 2} ]~,
\end{equation}
allowing the momentum equation to be written as,
\begin{equation}
\partial_t v_k +v_j \partial_j v_k +\partial_k 
[g\rho/m^2 -(\hbar^2/2m)\rho^{-{1 \over 2}} 
\partial_j \partial_j \rho^{1 \over 2}+V/m ]=0~.
\label{eq:euler}
\end{equation}
The conservation of energy (Bernoulli equation) then follows 
as an integral of Euler's equation,
or more directly from the real part of equation (\ref{eq:nlse1}):
\begin{equation}
\hbar \partial_t \phi+ {\textstyle{ 1 \over 2}}m v^2 + 
g\rho/m -(\hbar^2/2m)\rho^{-{1 \over 2}} \nabla^2 \rho^{1 \over 2}+V =0~.
\label{eq:bern}
\end{equation}
Perhaps the most significant quantum effect on the mechanics of the 
fluid is the quantisation of angular momentum.  
The circulation is given by,
\begin{equation}
\Gamma = \oint {\rm d}\mbox{\boldmath $r$} \cdot \mbox{\boldmath $v$} = (\hbar/m)2\pi s\ \ \ \ \ s=0,1,2, \dots 
\end{equation}
where the closed contour joins fluid particles. 
The conservation of angular momentum (Kelvin's theorem), follows from
Euler's equation (\ref{eq:euler}) and states that
the circulation around a closed `fluid' contour does not change in time. 
This means that within the fluid, vortex lines must created in pairs
which emerge from a point. The exception is at boundaries, where the wavefunction is clamped
to zero and no closed fluid loop can be drawn, e.g., at the surface
of an impenetrable object \cite{land87} 
or from the edge of a trapped condensate \cite{cara99}.

\subsection{Units}
For a homogeneous fluid flow, where the external potential is due to the obstacle only, 
it is convenient to rescale length and velocity in terms of the healing length $\xi=\hbar/\sqrt{mn_0g}$
and the asymptotic speed of sound $c=\sqrt{n_0g/m}$, respectively.
In this case, equation (\ref{eq:nlse1}) becomes
\begin{equation}
{\rm i} \partial_t \tilde{\psi}(\mbox{\boldmath$r$},t) = \left[ 
- \textstyle{1\over2} \nabla^2  
+ V(\mbox{\boldmath$r'$})
+\vert\tilde{\psi}(\mbox{\boldmath$r$},t)\vert^2\right] \tilde{\psi}(\mbox{\boldmath$r$},t)~,
\label{eq:nlse2}
\end{equation}
where $\tilde{\psi}=\psi/\sqrt{n_0}$ and $n_0$ is the number density far from the 
object. The force per unit length is measured in units of $\hbar\sqrt{n_0^3g/m}$. 
Unless otherwise stated we use these units throughout.
For steady flow, in which $\mbox{\boldmath $ v$}$ and $n$ are independent of time and 
$\phi=-\mu t$, where $\mu$ is the chemical potential, 
the Bernoulli equation takes the form
\begin{equation}
n-{\textstyle{1 \over 2}}
(\sqrt{n})^{-1}\nabla^2\sqrt{n}+
V+{\textstyle{ 1 \over 2}}v^2={\rm constant}~.
\label{eq:bern2}
\end{equation}

\section{The critical velocity}
\label{sec:vc}

The critical velocity for the breakdown of superfluidity is 
often expressed in terms of the Landau condition 
\cite{nozi90}, $v_{\rm c}=(\epsilon/p)_{\rm min}$,
where $\epsilon$ and $p$ are the energy and momentum of elementary
excitations in the fluid, and $v_c$ is the flow velocity in the
fluid bulk . In the dilute Bose gas, the long wavelength 
elementary excitations are sound waves and the Landau criterion 
predicts that $v_{\rm c}=c$. However, for flow past an object, 
the local velocity near the obstacle, $v$, can become supersonic even 
when flow velocity, $U$, is subsonic. Consequently, the critical flow
velocity, $v_{\rm c}$, where laminar flow becomes unstable occurs
at a fraction of the sound speed.
An estimate of $v_{\rm c}$ may be found following the argument suggested by 
Frisch {\it et al}. \cite{fris92}. For an incompressible flow past a 
solid object, Bernoulli's equation (\ref{eq:bern2}) (neglecting the 
quantum pressure) has the simple form,
\begin{equation}
n(v)+{\textstyle{1 \over 2}} v^2 = 1 +{\textstyle{1 \over 2}} U^2~,
\end{equation}
where $U$ is the background flow velocity.
The maximum velocity which occurs at the equator of the object is 
$v={\textstyle{3 \over 2}}U$ for a sphere (or $v=2U$ for a cylinder), therefore,
${\textstyle{1 \over 2}}( {\textstyle{9 \over 4}}-1)U^2 = 1- n(v)$.
The critical velocity is reached when the maximum speed, $v={\textstyle{3\over2}}U$, 
is equal to the `bulk' sound speed, $c=\sqrt{n(v)}$, which gives 
$v_{\rm c}=U =\sqrt{8/23} \approx 0.59 $ 
(or $\sqrt{2/11}\approx 0.44$ for a cylinder). However, for a compressible
fluid, the equatorial velocity is slightly larger due to pressure effects.
The first-order correction gives 
$v={\textstyle{3\over 2}} U+ {\textstyle{1\over 2}} U^3$ which reduces the
critical velocity to $v_{\rm c} \approx 0.53$. 
 
The exact value of $v_{\rm c}$ may be
found by solving the uniform flow equation \cite{huep97,wini99b}, 
\begin{equation}
{\rm i} \partial_t \tilde{\psi'}(\mbox{\boldmath$r'$},t)
=  \left[- \textstyle{1\over2} \nabla'^2  
+ V(\mbox{\boldmath$r'$})
 + \vert\tilde{\psi'}(\mbox{\boldmath$r'$},t) \vert^2 
+{\rm i}\mbox{\boldmath$v$} \cdot 
\nabla'\right]\tilde{\psi'}(\mbox{\boldmath$r'$},t)~,
\label{eq:uflow}
\end{equation}
where $\tilde{\psi'}(\mbox{\boldmath$r'$},t)=
\tilde{\psi}(\mbox{\boldmath$r$},t)$ is the wavefunction in the
fluid rest frame written
in terms of the object frame coordinates, $\mbox{\boldmath$r'$}=\mbox{\boldmath$r$}-\mbox{\boldmath$v$}t$. Stationary solutions of the form, $\tilde{\psi'}(\mbox{\boldmath$r'$},t)=
\phi(\mbox{\boldmath$r'$}){\rm e}^{-{\rm i}\mu t}$ are
found to exist only for $v\leq v_{\rm c}$, where $v_{\rm c}$ is the critical
velocity for vortex formation.

\begin{figure}
\begin{center}
\epsfig{file=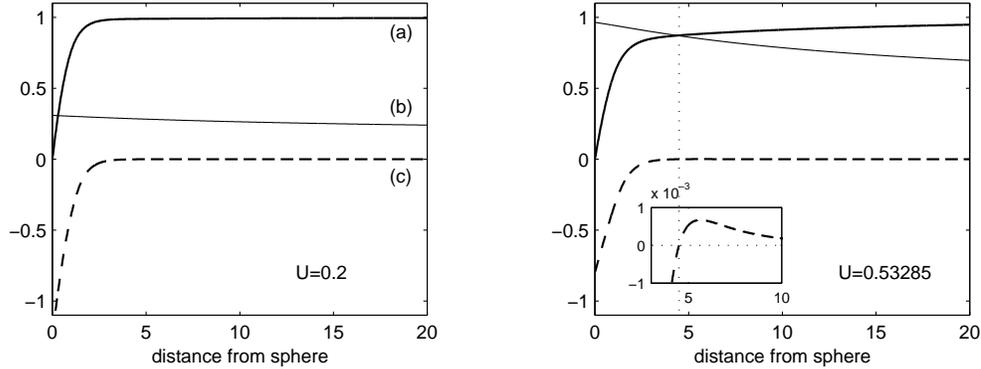,clip=,width=13cm,bbllx=75,bblly=400,bburx=540,bbury=580}
\end{center}
\caption{Laminar flow past a sphere with radius $R=50$ for two
flow velocities $U=0.2$ (left) and $U=0.53285$ (right). The three curves
show, (a) the `bulk' sound speed $c=\sqrt{n}=\vert\psi\vert$, 
(b) the velocity, $v$, and (c) the 
quantum pressure, $\nabla^2\psi/2\psi$,
as a function of position. Note that, at the critical velocity $u=0.53285$ (right),
the fluid velocity is equal to the sound speed where the quantum pressure
is exactly zero. 
}
\label{fig:1}
\end{figure}

To illustrate the behaviour of the exact solutions near the critical velocity,
we solve equation (\ref{eq:uflow}) in 3D for an impenetrable sphere
with radius $R=50$. The wavefunction, velocity and quantum pressure 
term near the object
are shown in Fig.~\ref{fig:2}. Note that these parameters are related via
the Bernoulli equation (\ref{eq:bern2}). 
The intersection between the velocity $v$ and wavefunction amplitude, $\vert\psi\vert$, 
curves defines the position where the velocity is equal to the `bulk' sound speed (\ref{eq:sound}). 
Note that close to the object the effective sound speed is increased
due to the quantum pressure term, (\ref{eq:pressure}),
therefore even though the density is low
the flow is not `supersonic'. The
critical velocity is reached when flow velocity exceeds
the speed of sound in the bulk of the fluid, i.e., when the intersection
between the velocity and wavefunction curves moves into the 
region where the quantum pressure term is zero, see Fig.~\ref{fig:1}(right).

The complication for trapped condensates is that the speed of sound and hence
the critical velocity depend upon position. In addition,
the object potential is typically penetrable and non-uniform. 
We return to these topics in Sec.~\ref{sec:trap}.

\section{Vortex shedding, drag, and dissipation}
\label{sec:drag}

\begin{figure}
\begin{center}
\epsfig{file=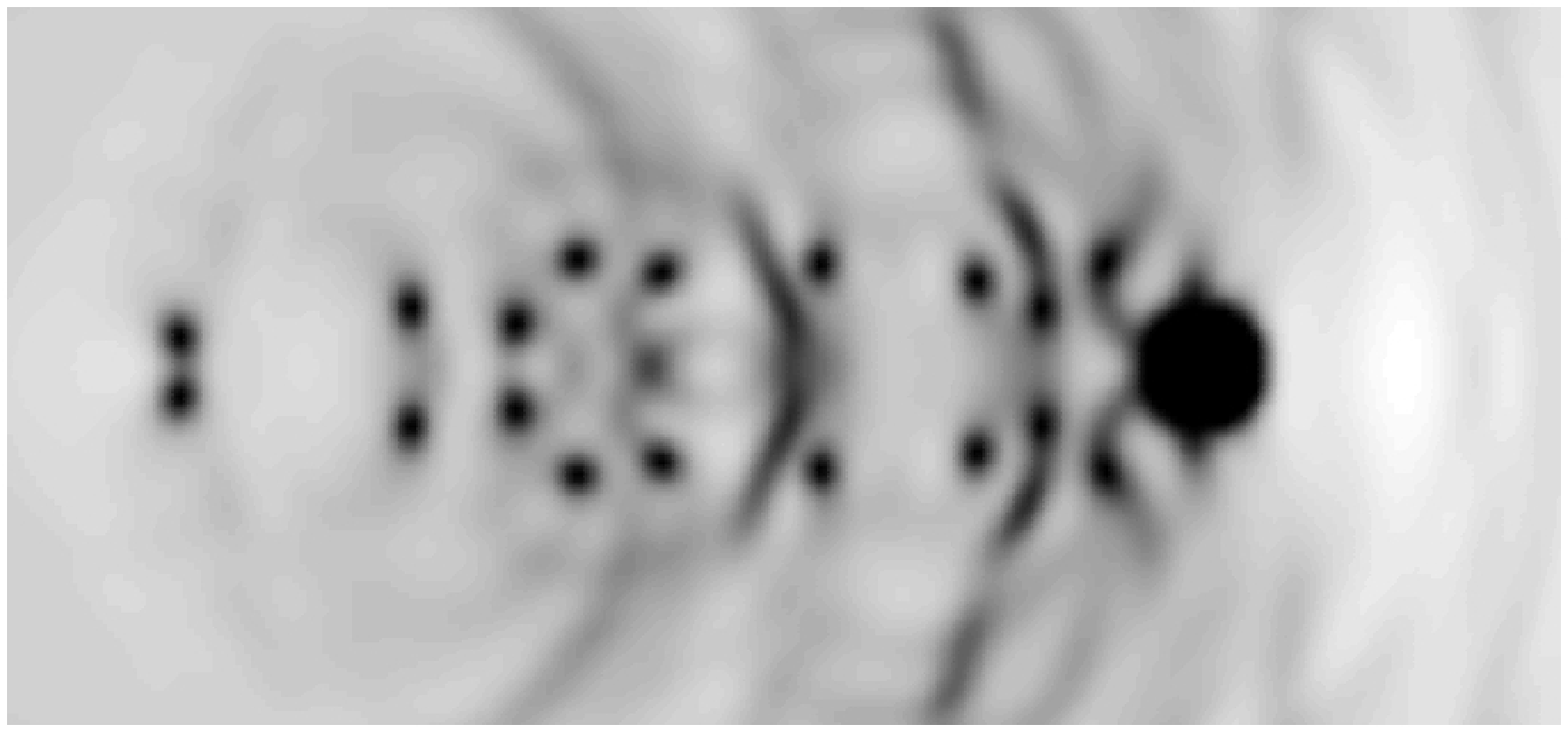,clip=,angle=-90,width=10cm,bbllx=165,bblly=170,bburx=415,bbury=695}
\end{center}
\caption{A shaded contour plot showing the condensate density
for homogeneous flow past a cylinder with radius $R=3$. The dark regions indicate lower
density, i.e., the position of the object and vortex lines.}
\label{fig:2}
\end{figure}

For flow faster than the critical velocity, $v>v_{\rm c}$, vortices are emitted
approximately periodically. A typical vortex stream pattern
for flow past an impenetrable cylinder is shown in Fig. \ref{fig:2}. 
The background flow is from right
to left and the vortex - anti-vortex pairs produce a flow pattern which 
opposes the background. Consequently, the
vortex trail separates the main flow from an almost stationary wake.
The momentum loss from the fluid is transferred to the object
creating a drag force. 
The contribution of vortex shedding to the drag force 
can be estimated by considering the momentum transfer
due to vortex emission, i.e., 
\begin{equation}
\mbox{\boldmath $F$}_{\rm v}=f_{\rm v}\mbox{\boldmath $p$}_{\rm v}~,
\label{eq:f_v}
\end{equation}
where $f_{\rm v}$ is the vortex shedding frequency 
and $\mbox{\boldmath $p$}_{\rm v}$ is the momentum of a vortex pair
as it is created at the equatorial plane.
Small fluctuations in the vortex shedding frequency
occur because as the vortices move
downstream they interact with each other creating fluctuations in the
flow pattern around the object. The drag force is taken to be
the time-average over many vortex emission cycles.

\begin{figure}
\begin{center}
\epsfig{file=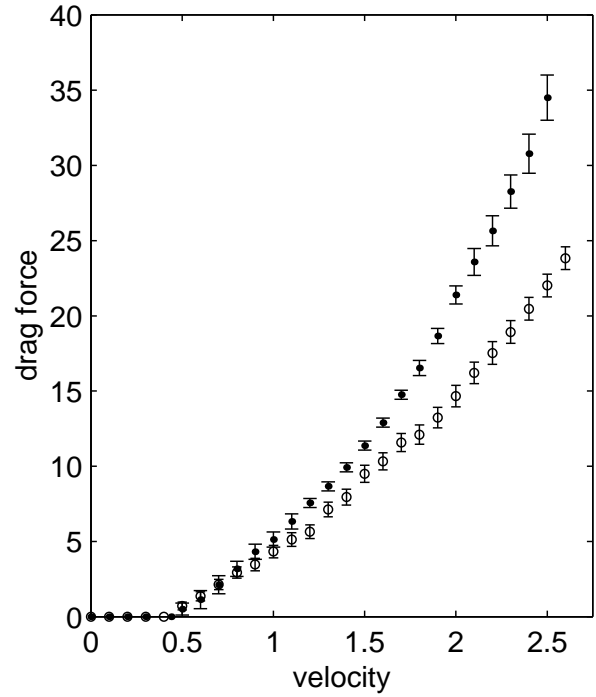,clip=,width=8cm,bbllx=50,bblly=180,bburx=255,bbury=420}
\end{center}
\caption{The time-averaged drag force as a function of velocity
for an impenetrable cylinder with radius $R=3$.
The open circles indicates the contribution due to vortex shedding. 
The error bars indicate the residual fluctuations in the time-averaged drag.}
\label{fig:3}
\end{figure}

\begin{figure}
\begin{center}
\epsfig{file=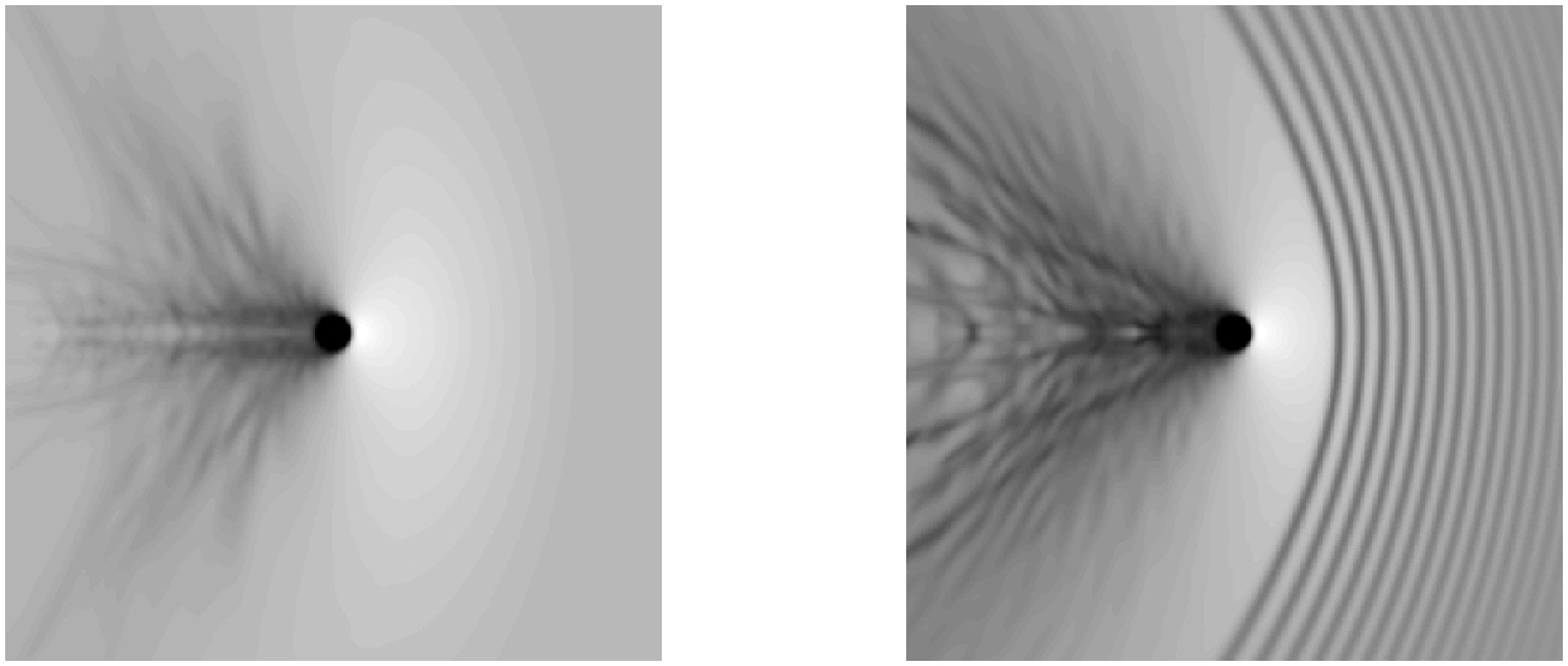,clip=,angle=-90,width=13cm,bbllx=130,bblly=75,bburx=430,bbury=780}
\end{center}
\caption{Time-averaged density for homogeneous flow past a cylinder  
(radius $R=3$) for flow speeds of 0.9 (left) and 1.2 (right) times the speed of
sound. For supersonic flow a standing wave pattern appears in front of the object.}
\label{fig:4}
\end{figure}

In addition to vortex shedding, drag may also arise due to sound waves.
The time-average drag, evaluated using equation (\ref{eq:force}), as a function of velocity
for an impenetrable cylinder is shown in Fig.~\ref{fig:3}.
The drag is zero up to the critical velocity, then increases 
approximately quadratically with $v$ \cite{fris92,wini99a}.
Also shown is the contribution to the drag force from vortex
shedding alone, i.e., equation (\ref{eq:f_v}).  
This comparison illustrates that for $v<c$ the drag is produced by vortex
shedding, whereas for $v>c$, an increasingly significant contribution arises from
sound waves. For $v>c$, the 
reflected matter waves create a standing wave pattern 
in front of the object as shown in Fig.~\ref{fig:4}.

Using similar arguments for energy loss of the flow. Consider now
the condensate at rest and the obstacle moving with speed $v$. 
A drag force leads to energy transfer to the condensate. The
energy transfer rate is given by
\begin{equation}
\frac{{\rm d}E}{{\rm d}t}=\mbox{\boldmath $F$}_{\rm drag} .\mbox{\boldmath $v$}~.
\label{eq:diss}
\end{equation}
Eqs.~(\ref{eq:f_v}) and (\ref{eq:diss}) make the important link between
vortex shedding, drag and energy dissipation.

\section{Motion in a trapped condensate}
\label{sec:trap}

The inhomogeneous density profile and finite size of trapped condensates means that
a steady, uniform flow is difficult to achieve. 
The MIT experiment \cite{rama99} partially overcame
this problem by sweeping the object back and forth at constant velocity
within the central region of the
condensate, where the density is approximately uniform. In this
case, the object moves through its own low-density wake, 
and consequently the drag law
is different from the uniform flow case discussed above.

\begin{figure}
\begin{center}
\epsfig{file=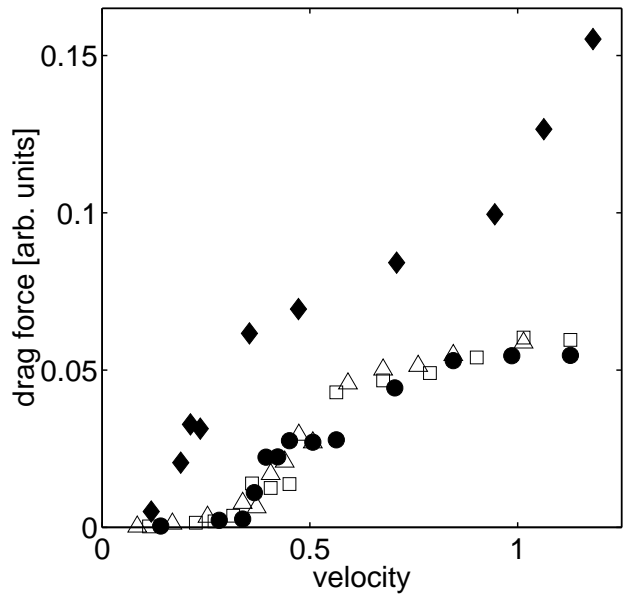,clip=,width=10cm,bbllx=35,bblly=180,bburx=545,bbury=585}
\end{center}
\caption{The drag force as a function of velocity for a laser beam
moving back and forth through an inhomogeneous condensate. The triangles, squares
and circles corresponds to 2D simulations with different oscillation amplitudes.
The diamonds correspond to a 3D simulation. In 2D the force is effectively that on 
a disk of depth 2.5 healing lengths, whereas in 3D the condensate depth corresponds
to a Thomas Fermi radius of 6.4 healing lengths.}
\label{fig:5}
\end{figure}

In Fig.~\ref{fig:5} we show the time-averaged drag on a laser beam oscillating
in a trapped condensate. Important differences with the uniform flow case
(Fig.~\ref{fig:3}) arise at high and low velocities. The drag force tends to saturate
at higher velocities as the object expels fluid from the region of
oscillation and the pressure drops.

Fig.~\ref{fig:5} also highlights important differences between two and three
dimensions. Two dimensions corresponds to the limit of a `cylindrical'
condensate, where the density is uniform parallel to the axis of the object
(which we define as the $z$-axis). However, in realistic three-dimensional
situations the density is inhomogeneous along $z$, leading to a variation of
the speed of sound which vanishes at the condensate edge. This results in a
lower critical velocity in 3D than in 2D, as apparent in Fig.~\ref{fig:5}. 
In the Thomas-Fermi limit ($ga^3\gg 1 $) the density profile is parabolic, therefore 
the average sound speed along $z$ is a factor of $\pi/4$
smaller than that at the centre. The remaining reduction arises from the fact that 
even at very low velocities, vortices tend to be formed where the laser beam intersects the 
lower density fluid at the condensate edge. 
The lower densities arising in 3D also leads to enhanced sound emission, and hence
enhanced drag at high velocities.

\begin{figure}
\begin{center}
\epsfig{file=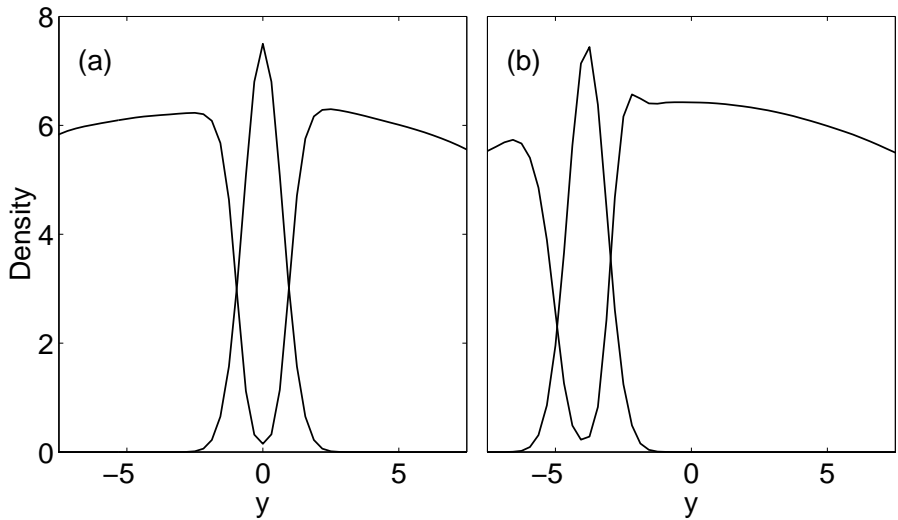,clip=,width=12cm,bbllx=70,bblly=205,bburx=580,bbury=500}
\end{center}
\caption{Cross-section through a 2D trapped condensate pierced by a Gaussian
laser beam `object' (also shown). (a) If the object is moving at constant velocity, 
the fluid density is symmetric around the object, and the net drag is zero. (b) 
If the object accelerates the fluid cannot adjust sufficiently rapidly, leading to an asymmetric fluid
distribution around the object which produces drag.
The fluid relaxes by the emission of a sound wave. The $x$-axis is in units of the laser
beam waist, while for the $y$-axis the units are arbitrary.}
\label{fig:6}
\end{figure}

Below the critical velocity, dissipation due to sound emission 
occurs at the motion extrema, where the 
object accelerates. This is illustrated
in Fig.~\ref{fig:6} which shows a cross section through a 2D condensate cut
by a moving laser beam. For constant motion, Fig.~\ref{fig:6}(a), the fluid is distributed 
symmetrically around the object and the drag, which is given by an overlap
integral between the condensate density and the gradient of the object potential (i.e., the second term in equation~(\ref{eq:force})) is zero. 
When the object accelerates, Fig.~\ref{fig:6}(b), the fluid fails to respond 
rapidly enough to the abrupt change in velocity, and
the asymmetry in the overlap between the fluid and the objects leads to 
a resistance force analogous to dynamic buoyancy. The system
relaxes to the uniform flow case by the emission of a sound wave. 
This corresponds to the `phonon heating' process discussed
in \cite{jack00b}.

Although the NLSE model describes dissipation in the sense of
energy transfer between a moving object and the fluid, it
does not say anything about how that energy (mostly stored within the vortex core) 
may be subsequently converted into heat.
A complete description should include
coupling of the condensate to a thermal cloud, and would describe the damping
of phonon and vortex modes. Recent work on the non-equilibrium dynamics of 
the condensate and non-condensate predict a depletion of the 
condensate fraction \cite{niku99} as observed in the MIT experiment.

\section{Summary}

The motion of an object through a dilute Bose-Einstein condensate
provides an ideal system to study the fundamental
problem of the onset of dissipation in superfluids. 
In this paper we have explained the role of vortex shedding and sound emission
in energy transfer between the object and the
condensate. No energy transfer is observed under the condition of uniform, 
steady flow at
speeds below a critical velocity. However, if the object accelerates
there is a small dissipative effect, even below the critical
velocity, due to sound emission. The critical velocity is reached
when the local velocity in the bulk of the
fluid (i.e., where the quantum pressure is zero) exceeds the speed of
sound. 
Above the critical velocity vortices are emitted leading to a drag force and
energy transfer to the fluid. Vortex shedding
dominates the energy transfer for intermediate velocities, while sound
emission becomes increasingly important for supersonic motion. 
We highlight some important differences
between homogeneous and inhomogeneous trapped condensates. 
In particular, that the critical velocity is substantially reduced when the object 
intersects regions of lower density at the condensate edge and that
the trap inhomogeneity gives rise to an additional term in 
the drag force analogous to a buoyancy. 

\ack This work is supported by the EPSRC.


\section*{References}

\begin{thebibliography}{99}

\bibitem{fermi} See e.g. {\it Bose-Einstein condensation in atomic gases}, Proc. 
Int. School of Physics Enrico Fermi, eds. M. Inguscio, S. Stringari and C.
Wieman (IOS Press, Amsterdam, 1999).

\bibitem{madi00} K. W. Madison, F. Chevy, W. Wohlleben, and J. Dalibard,
Phys. Rev. Lett. {\bf 84}, 806 (2000).

\bibitem{matt99} M. R. Matthews, B. P. Anderson, P. C. Haljan, D. S. Hall, 
C. E. Wieman, and E. A. Cornell, Phys. Rev. Lett. {\bf 83}, 2498 (1999).

\bibitem{donn91} R. J. Donnelly {\it Quantized vortices in Helium II}, (CUP,
Cambridge, 1991).

\bibitem{rama99} C. Raman, M. K\"ohl, R. Onofrio, D. S.
Durfee, C. E. Kuklewicz,  Z. Hadzibabic, and W. Ketterle, Phys. Rev. Lett.
{\bf 83}, 2502 (1999).

\bibitem{ginz58}
V. L. Ginzburg and L. P. Pitaevskii, Sov. Phys. JETP {\bf 7}, 858 (1958); E.
P. Gross, J. Math. Phys. {\bf 4}, 195 (1963).

\bibitem{fris92} T. Frisch, Y. Pomeau, and S. Rica, Phys. Rev. Lett. {\bf 69}, 
1644 (1992).

\bibitem{jack00b}
B. Jackson, J. F. McCann, and C. S. Adams, Phys. Rev. A {\bf 61}, 051603 (R)
(2000).

\bibitem{nozi90} P. Nozi\`eres and D. Pines, {\it Theory of Quantum Liquids Vol II}
(Addison-Wesley, Redwood City, 1990).

\bibitem{land87}
{\it Fluid Mechanics (2nd ed.)}, L. D. Landau and E. M. Lifshitz (Pergamon,
 Oxford, 1987).

\bibitem{wini99a}
T. Winiecki, J. F. McCann, and C. S. Adams, Phys. Rev. Lett. {\bf 82}, 5186 (1999).

\bibitem{jack98}
B. Jackson, J. F. McCann, and C. S. Adams, Phys. Rev. Lett. {\bf 80}, 3903
(1998).

\bibitem{cara99}
 B. M. Caradoc-Davies, R. J. Ballagh, and K. Burnett, Phys. Rev. Lett.
 {\bf 83}, 895 (1999).

\bibitem{huep97}
C. Huepe and M.-\'E. Brachet, C. R. Acad. Sci. Paris, {\bf 325}, 195 (1997).

\bibitem{wini99b} T. Winiecki, J. F. McCann, and C. S. Adams, 
Europhys. Lett. {\bf 48}, 475 (1999).

\bibitem{niku99}
 T. Nikuni, E. Zaremba, and A. Griffin, Phys. Rev. Lett. {\bf 83}, 10 (1999).

\end{thebibliography}
\end{document}